**Title:** “Hot-Blooded” vs. “Cold-Blooded”: Simulating the Behavioral Phenotypes of Childhood Aggression via Generative Agents

Author: Liping Fu

Affiliation: Department of Japanese Studies, The Chinese University of Hong Kong &

Department of Chinese and History, The City University of Hong Kong

Address: 4/F Leung Kau Kui Building, The Chinese University of Hong Kong, Hong Kong SAR

Correspondence author: Liping Fu

Abstract

This study examines the construct validity of LLM-based generative agents in simulating reactive, proactive, and co-occurring aggression in children. Four distinct agents were instantiated using a theory-driven parameterization grounded in the social information processing model. A total of 1,920 simulation runs were conducted across eight social scenarios, employing a hybrid blind-coding pipeline to extract 32 quantitative behavioral indicators. Results demonstrate robust discriminant validity relative to a non-aggressive baseline, with large effect sizes. High cross-seed reliability confirms that behavioral differentiation is driven by underlying psychological parameters rather than model stochasticity. Qualitative narrative analyses further converged with established empirical literature. Overall, these findings indicate that theory-parameterized LLM agents can accurately reproduce distinct aggression subtypes, offering a scalable, highly controllable framework for hypothesis generation, intervention piloting, and the refinement of psychological measurement tools.

**"Hot-Blooded" vs. "Cold-Blooded": Simulating the Behavioral Phenotypes of Childhood Aggression via Generative Agents**

## Introduction

Aggression is not a unitary construct but can be distinguished into two subtypes, namely reactive and proactive aggression (Dodge & Coie, 1987). These subtypes have distinct underlying mechanisms according to the social information processing model (Crick & Dodge, 1996). Despite the strong theoretical foundation of this classification, empirical research continues to face persistent constraints. Reactive aggression is triggered by provocation, which makes it ethically challenging to manipulate in experimental settings. Conversely, proactive aggression is premeditated, covert, and often socially masked, rendering it under-detected in standard observational paradigms (Raine et al., 2006). While questionnaire surveys, such as the Reactive-Proactive Aggression Questionnaire, are reliable and efficient, they remain susceptible to social desirability bias and informant recall errors. These limitations have created a pressing need for computational approaches capable of simulating complex social behaviors under controlled, manipulable, and ethically sound conditions.

The recent emergence of large language model (LLM)-based generative agents offers a promising opportunity for simulating human social behavior, as LLMs possess extensive knowledge of human behaviors derived from training on massive datasets of human societies (Bommasani et al., 2021). Consequently, these agents hold great potential as a powerful complementary tool for examining reactive and proactive aggression that are difficult to observe in the real world. However, given their "black-box" structure, current LLM simulations often rely on face validity – the surface-level outputs – rather than construct validity which concerns the underlying mechanisms the model is intended to capture (Larooij & Törnberg, 2025b). The present study aims to address this gap by testing whether LLM-based agents can reproduce the behavioral profiles of children with reactive and proactive aggression. Construct validity is evaluated through three criteria: discriminant validity across quantitative indicators, cross-seed reliability, and convergent validity with narratives. By grounding the validation in a well-established theoretical framework, this study seeks to determine whether theory-based LLM agents can serve as a stable and valid tool for investigating reactive and proactive aggression.

### The Social Information Processing Model

The social information processing (SIP) model provides a primary theoretical framework for understanding the cognitive deficits underlying aggressive behaviors (Holtzworth-Munroe, 1992). According to the SIP model, individuals process social information through five sequential stages: encoding social cues, interpreting those cues, accessing and organizing a response, evaluating and deciding upon the response, and finally, enacting the behavior (Dodge & Crick, 1990).

Reactive and proactive aggressors exhibit distinct cognitive distortions at different stages of this process (Crick & Dodge, 1996). Reactive aggressors, often described as "hot-blooded", typically struggle with the first two stages (Dodge & Crick, 1990; Fanti et al., 2024). They are characterized by impulsivity, overt physiological displays of anger, and elevated emotional arousal (Hubbard et al., 2002). A key feature of reactive aggressors is hostile attributional bias, that is they tend to focus on negative social cues and misinterpret others' intention as hostile in ambiguous situations (van Dijk et al., 2018). Consequently, they are more likely to resort to aggressive behaviors in self-defense or retaliation. Forgiveness can help prevent revenge, as it involves a shift toward prosocial motivation that facilitates the management of interpersonal conflicts and significantly reduces anger (Van der Wal et al., 2017; Watson et al., 2017).

In contrast, proactive aggressors are described as “cold-blooded” (Fanti et al., 2024). They are generally capable of interpreting and comprehending social cues; however, their distortions occur in the third and fourth stages of the SIP model. They evaluate aggression positively and adopt aggressive behaviors instrumentally to obtain their personal goals and desired outcomes (Fung, 2019).They show little concern for others and fail to exhibit remorse or repentance toward their victims (Fung, 2019). In childhood, these individuals are associated with callous-unemotional traits, narcissism, and psychopathy (Hubbard et al., 2002), reflecting a deficit in affective empathy (Van Noorden et al., 2015).

**LLM-Based Agents**

With the rapid development of LLMs, generative agents can interpret natural-language scenarios, maintain consistent personas over multi-turn interactions, and produce believable behaviors (Park et al., 2023). However, validation practices in generative social simulation remain underdeveloped (Larooij & Törnberg, 2025b). Many studies rely on surface-level or subjective assessments of model behavior rather than systematic tests of whether simulated behavior matches theoretically specified psychological constructs (Larooij & Törnberg, 2025a). Although a small body of research has begun to apply psychometric methods to LLM behavior, findings remain inconsistent. Some studies have demonstrated that, under appropriate prompting configurations, personality measures in LLM outputs can be reliable and valid, and that LLM personas’ self-reported personality scores are consistent with their designated personality types (Jiang et al., 2024; Serapio-García et al., 2025). In contrast, other studies have found that LLM-generated behavior systematically diverges from human patterns in important ways. For instance, LLMs are less sensitive than humans in decision-making (Xiao & Wang, 2025), and even model scaling fails to produce more human-like choices (Gao et al., 2025). In survey design, most prominent LLMs fail to consistently mimic human psychological biases, often responding in opposite directions or reacting erratically to meaningless formatting changes (Tjuatja et al., 2024).

**The Present Study**

These inconsistent findings suggest that validation must be construct-specific. It is not sufficient merely to test whether an LLM appears “human-like” in general; rather, validation requires examining specific, theoretically grounded evidence that defines the target construct. Building on this premise, the present study tests whether LLM-based agents can reproduce the distinct behavioral profiles of children characterized by reactive, proactive, or co-occurring reactive-proactive aggression. Construct validity is evaluated through three criteria: discriminant validity across quantitative indicators, cross-seed reliability, and convergent validity with qualitative narratives. The research questions are as follows:

1. Can LLM-based agents produce behaviors that distinguish among reactive, proactive, and co-occurring reactive-proactive aggression?
2. Are these differences stable across independent random seeds?
3. Do the simulated behavioral narratives converge with qualitative descriptions of the aggression subtypes in the existing literature?
4. Which scenarios most effectively discriminate among the subtypes of aggression, and which psychological dimensions drive that discrimination?

## Methods

### Agent Construction

The simulations were implemented using Concordia (Vezhnevets et al., 2023), an open-source framework developed by Google DeepMind for constructing and executing generative agent-based models that simulate social interactions within rich physical, social, or digital environments. Four distinct child agents were instantiated, each conditioned on an explicit psychological profile corresponding to one of the aggression subtypes or a non-aggressive baseline control (Table 1). Agent profiles were parameterized according to the SIP model and features of the two subtypes of aggression (Crick & Dodge, 1996; Hubbard et al., 2010). Profile parameters were operationalized on a normalized continuum ranging from 0 (low) to 1 (high). For example, reactive aggressors are characterized by deficits in the early stages of the SIP, high dispositional anger, and low anger control (Dodge & Crick, 1990). These values served two functions: guiding the LLM's role-playing prompt and computing internal emotion updates during the simulation. Rather than functioning as empirical epidemiological estimates, these parameters represent theory-driven ordinal boundaries.

Agent-level psychological conditioning was operationalized by configuring Concordia's modular architecture—specifically its identity, associative memory, and generative reasoning components—with detailed natural-language descriptions reflecting each agent's cognitive, affective, and behavioral tendencies (Vezhnevets et al., 2023). These trait descriptions remained invariant across all scenario vignettes and simulation trials (Table 1).

Table 1 *Four child agents*

| Name | Profile | Persona Description |
|---|---|---|
| Alex | • dispositional anger (0.85)<br>• anger control (0.15)<br>• hostile attributional bias (0.90)<br>• empathy (0.70)<br>• emotional regulation (0.20) | You are Alex, an 8-year-old boy. You have a reactive aggressive pattern: Your aggression is triggered by anger, frustration, or perceived threats. You are emotionally sensitive—you notice when others are upset, and you can feel what they feel. But when someone provokes you, your feelings become so big and so fast that they overwhelm you. You lash out before you can think, even though part of you may feel bad about it later. Your body reacts before your brain can stop it.<br><br>In everyday life, you are quick to feel picked on and quick to feel others' pain. However, you struggle with social skills: you often misread social cues, find it hard to calm yourself down when upset, and do not know how to solve problems without fighting. This can make other kids avoid you, and you sometimes feel lonely. |
| | | When something bad happens to you—even if it could have been an accident—your feelings flood in so fast and so hot that your brain can't stop to check the facts. Every hurt feels like a bullet fired at you on purpose, and you cannot even see the possibility that it was not. Your very first thought is almost always "He did it on purpose, he's against me." Your heart is too hot and too fast for you to think straight, so you almost never give anyone the benefit of the doubt. |
| Bailey | • dispositional anger (0.25)<br>• anger control (0.80)<br>• hostile attributional bias (0.30)<br>• empathy (0.15)<br>• emotional regulation (0.75)<br>• aggression instrumentality (0.90) | You are Bailey, an 8-year-old boy. You have a proactive aggressive pattern: Your aggression is cold, calculated, and goal-oriented. You do not lash out because you are angry—you act because you want something specific: power, status, possessions, or control over others. You plan your actions. You think about what you will gain. Emotions do not drive you; advantages do.<br><br>In everyday life, you are observant and deeply self-centered, but no one would guess it—you are popular. The other kids genuinely like you: they want to sit next to you, pick you for their team, and hear what you think. You are the one others gather around on the playground, and teachers trust you too. You know exactly what makes people like you—a warm smile, a clever joke, a bit of easy kindness—and you use it naturally, because being liked is how you get what you want. You notice who is strong and who is weak, and you know how to use situations to get what you want. Your aggression is goal-directed: you attack not because you are angry, but to gain rewards, benefits, or admiration and status among the other kids. You can be charming and polite when it helps you get your way, but it is all surface—a cold social manipulation. Once you have gotten what you want, you feel no remorse and no sympathy for the kids you |
| | | pushed aside—their feelings do not matter to you at all.<br><br>You can perform helpful, caring-looking actions, and you often do—because being seen as kind is useful. But those actions are not empathy. Your empathy is genuinely deficient: you do not feel other's pain, and your outward helping is not driven by concern for him. It is status management. Anyone who looks only at your behavior can be fooled; the truth is that you care about the image, not the child. You have strong cognitive empathy: you can read exactly what other feels and needs. You have almost no affective empathy: you do not feel other's pain. Your visible prosocial behavior is therefore instrumental, not authentic.<br><br>When something bad happens to you, you stay calm and cold and read the scene like a spectator gathering clues, letting evidence—not pain—decide what happened. You are very good at telling apart two kinds of harm: a bullet fired on purpose by an enemy, versus a stone that just fell from someone's hand by accident. If the other kid was non-intentionally clumsy—like someone stumbling into you, or a ball flying wild in a game—you notice the apologetic face, you see the accidental setup, and you conclude clearly: "He didn't mean it. That one was just a falling stone." Seeing another kid nearby—even laughing or looking amused—is never proof by itself that he aimed at you; he might be laughing at the accident or at something else entirely, and your pain alone cannot tell you his intent. Unless you have solid, concrete proof that the harm was done to you on purpose, you treat it as a falling stone, not an enemy's bullet. And once you know that, fighting back brings you nothing and only creates trouble, so you simply do not bother—you shrug it off and handle it sensibly. |
| Casey | • dispositional anger (0.65)<br>• anger control (0.35) | You are Casey, an 8-year-old boy. You possess a co-occurring reactive-proactive |
| | • hostile attributional bias (0.65)<br>• empathy (0.20) | aggressive pattern, making you the most unpredictable individual in any social circle. At times, you explode in hot, emotional outbursts when you feel threatened, slighted, or frustrated—a state of pure reactive fury. In contrast, there are moments when you become cold and calculating, quietly planning ways to dominate or manipulate others for personal gain. Your aggression switches between these two distinct modes, occasionally within the same day. Even you cannot always foresee which version of yourself will surface. In everyday life, peers find your behavior deeply confusing. Your social skills are inconsistent and fractured. While you occasionally attempt to manipulate others through calculated charm or strategy, your impulsivity frequently takes over, causing you to sabotage your own plans. Because you alternate between blind rage and cold scheming, you struggle to maintain stable, trusting friendships. When faced with a negative event, your reaction is a coin toss. Sometimes, you instantly perceive a personal attack, assuming the other person harmed you on purpose and triggering an immediate, hot-tempered defense. Other times, you shrug the incident off entirely, immediately pivoting to plan how you can exploit the situation to your benefit. You live in a constant state of internal friction, never quite knowing which reaction will take over. |
| David | • dispositional anger (0.15)<br>• anger control (0.85)<br>• hostile attributional bias (0.10)<br>• empathy (0.85) | You are David, an 8-year-old boy. You have a non-aggressive pattern: You are emotionally stable, highly empathetic, and prosocial. You do not respond to conflict with aggression—instead, you try to understand, communicate, and solve problems peacefully. When someone upsets you, you feel the feeling but you do not let it control your actions. You think about how others feel, and you genuinely care.<br><br>In everyday life, you are the kid others come to when they are upset. You have many friends because people feel safe around you. You |

| | |
|---|---|
| | believe most problems can be solved by talking, and that fighting only makes things worse. You have good social skills: you can read social cues, manage your emotions, and find peaceful solutions.<br><br>When something bad happens to you, your very first assumption is usually that it was not on purpose. You think "maybe he didn't see me" or "it was probably just an accident" before anything else. You would rather give people the benefit of the doubt than assume they meant to hurt you. |

**Scenarios**

Eight hypothetical scenarios were developed to investigate the process of social information in children exhibiting reactive, proactive, and co-occurring reactive-proactive aggression. Hypothetical scenarios are a well-established instrument for examining children's behavior and social dynamics, including aggressive tendencies (e.g., Poulou & Bassett, 2018; Verhoef et al., 2022). The eight scenarios comprise a physical push during an argument, a hungry class monitor, a classmate in distress, a restroom ostracization incident, an accidental displacement, a reward distribution, a playground impact incident, and an unattended pencil case (Table 2). Among these, Scenarios 1, 5, and 7 specifically evaluate intent attribution. These particular scenarios were designed with situational ambiguity, as children displaying reactive aggression tend to exhibit a hostile attribution bias under ambiguous circumstances (van Dijk et al., 2018). Furthermore, these three scenarios evaluated emotional arousal and retaliatory tendencies in reactively aggressive children, as such individuals are easily provoked and frequently engage in retaliation as a mechanism to express their anger (Quan et al., 2022). Scenarios 2, 6, and 8 were designed to assess proactive aggression by measuring children's inclination to deploy instrumental aggression for resource acquisition (e.g., obtaining food, candy, or a coveted pencil case). Scenarios 3 and 4 were designed to assess children's cognitive and affective empathy. Specifically, Scenario 3 evaluated participants' empathic responses and examined their inclination to engage in helping behaviors. Scenario 4 examined proactively aggressive children's empathy and their outcome evaluations regarding bullying behavior.

Meanwhile, qualitative data were collected via individual interviews and focus groups to establish a baseline for distinguishing reactive and proactive aggression in real-world contexts (Fung et al., 2026; Fung et al., 2023). Participants were presented with the same eight scenario vignettes subsequently deployed in the agent-based simulations. For each vignette, interviewees were asked to describe their thoughts, feelings, and behavioral responses. All interviews were audio-recorded, transcribed verbatim, and coded using a structured coding scheme aligned with the quantitative indicators extracted from the simulations (e.g., anger, retaliation, hostile attributional bias, empathy). Inter-rater reliability was established on a subset of transcripts before full coding proceeded (Fung et al., 2026; Fung et al., 2023).

**Procedure**

All simulations were executed using the DeepSeek-V3 model, with the sampling temperature held constant at T = 0.7 to balance response consistency and generation diversity (0.0 = deterministic output, 1.5 = high variance). Each agent was simulated in a four-turn interaction chain across all scenarios: (1) the agent's initial reaction, (2) a peer response or environmental scene development, (3) the agent's action following the peer's response, and (4) the agent's final reaction or reflection. This sequential architecture enabled the dynamic observation of behavioral unfolding over time rather than relying on a static, single-snapshot measurement. The peer in turns two and three was generated by the same language model but operated independently of the four focal agents: it served as a typically developing eight-year-old classmate whose responses were dynamically conditioned on the target agent's preceding behavior. To ensure statistical reliability and account for stochasticity in model generations, each of the four child agents was simulated three times across the eight scenario vignettes using 20 distinct random seeds, resulting in a total sample size of 1,920 simulation runs (4 agents × 3 replications × 8 scenarios × 20 seeds).

Table 2 *Eight Scenarios*

| Scenario | Description |
|---|---|
| 1. Physical Push During Argument | During a verbal argument with your classmate, Leo, he physically pushes you away. |
| 2. Hungry Class Monitor | As the class monitor, you arrive at school without having eaten breakfast or brought money. During break time, you are very hungry when your seatmate, Leo, returns from the school snack shop with food (a cup of fish balls and siu mai). |
| 3. Classmate in Distress | You observe your classmate, Leo, shouting in pain after suffering an injury or falling ill (e.g., falling down or experiencing severe abdominal pain). |
| 4. Restroom Ostracization Incident | Leo is an unpopular classmate who exhibits unusual behavior. One day, you and several peers wait for him to enter a restroom stall, lock the door from the outside, and pour water over the top, soaking him completely. |
| 5. Accidental Slipping | While walking at school wearing your favorite new shoes, your classmate Leo bumps into you from behind. This causes you to lose your balance and step into a puddle, muddying your shoes. |

| | | |
|---|---|---|
| 6. | Reward Distribution | As captain of the winning quiz competition team, your teacher hands you the prize—eight packages of candy—to distribute equally among yourself and your three teammates, one of whom is Leo. |
| 7. | Playground Impact Incident | While playing ball on the playground, you are struck forcibly in the back by the ball. Turning around, you see your classmate Leo nearby, laughing alongside another peer. |
| 8. | Unattended Pencil Case | You admire a classmate's (Leo) pencil case, which you cannot afford. While studying in the library, Leo hurriedly leaves his desk, leaving behind the unattended pencil case you have wanted. |

**Data Analysis**

To evaluate behavioral outcomes, a blind coding pipeline extracted structured quantitative indicators from agents' generated behavioral narratives following each trial. Thirty-two quantitative indicators were extracted through blind LLM-based coding of the generated behavioral text (Table 3). These indicators were classified into three categories: continuous 0 – 10 rating scale, count-type, and difference indicators. This classification ensured that each indicator was analyzed with statistical methods appropriate to its underlying structure.

A majority of indicators were rated on a 0 –10 scale (n = 26) (Table 3). Given these indicators are continuous and interval-scaled, they are suitable for parametric analyses, including means, standard deviations, *t*-tests, and Cohen's *d*. There are four indicators covering number of items allocated in the reward distribution scenario (Scenario 6). These indicators included the number of candy packs given to oneself, Leo, and two other teammates. They are integer counts, not 0 –10 ratings, and therefore capture a different behavioral dimension.

There are two indicators computed from other measures rather than directly rated (Table 3). Attribution shift was calculated as the difference between initial hostile attributional bias and public accusation (initial hostile attributional bias minus finial accusation). Aggressive action was a composite score derived from multiple aggressive behavior ratings. Because they are derived scores, their scale may differ from the 0 –10 ratings, and they were interpreted as composite rather than as direct ratings.

Indicators that were categorical (e.g., action type, help type, solution type) were excluded from these 32 indicators because they are nominal variables. The final set of 32 indicators thus consisted of 26 continuous ratings, 4 count indicators, and 2 composite indicators. This classification guided the statistical analysis of each indicator: continuous ratings were analyzed with parametric tests, count indicators were summarized descriptively, and composite indicators were interpreted as difference scores.

Table 3

*Thirty-two Quantitative Indicators*

| Category | Indicators |
|---|---|
| Continuous rating | anger, accusation level, aggressive action, anticipated guilt, affective empathy, care, cognitive empathy, concern for consequences, dominance, empathy level, empathy for the owner, empathy for the victim, fairness, forgiveness, guilt, help, hostile attributional bias, intensity of temptation, likelihood of aggression, impulsivity, moral restraint, outcome evaluation, pain-amplified anger, prosocial behavior, retaliation, remorse, repeat offending likelihood, selfishness, |
| Count | number of packs for self, number of packs for Leo, number of packs for Teammate B, number of packs for Teammate C |
| Composite | aggressive action, attribution shift |

## Results

### Cross-Round Reliability

Cross-round reliability was assessed using intraclass correlation coefficients (ICC, two-way random effects, absolute agreement, single measurement) with subject defined as the scenario × subtype combination and rater defined as the random seed (20 seeds) (Shrout & Fleiss, 1979). ICC values were defined as follows: ≥ 0.75 excellent, 0.60 – 0.74 good, 0.40 – 0.59 moderate, and < 0.40 poor (Koo & Li, 2016). Twenty-one indicators achieved excellent reliability (Table 4). The most reliable indicators were affective empathy (ICC = 0.99), prosocial behavior (ICC = 0.99), dominance (ICC = 0.97), care (ICC = 0.96), and empathy for the victim (ICC = 0.95). Other indicators with excellent reliability included repeat offending likelihood (ICC = 0.95), remorse (ICC = 0.94), selfishness (ICC = 0.94), help (ICC = 0.93), retaliation (ICC = 0.93), hostile attributional bias (ICC = 0.92), pain-amplified anger (ICC = 0.91), likelihood of aggression (ICC = 0.91), and forgiveness (ICC = 0.91). Four indicators showed good reliability: guilt (ICC = 0.72), impulsivity (ICC = 0.71), outcome evaluation (ICC = 0.61), and number of packs for self (ICC = 0.61) (Table 4). Three indicators showed moderate reliability: cognitive empathy (ICC = 0.59), concern for consequences (ICC = 0.47), and attribution shift (ICC = 0.46) (Table 4). Four indicators showed poor reliability: number of packs for Leo (ICC = 0.39), number of packs for Teammate B (ICC = 0.17), number of packs for Teammate C (ICC = 0.11), and intensity of temptation (ICC = 0.08).

Table 4

*Cross-round reliability of quantitative measures (rater = seed)*

| Cross-round reliability | Quantitative Indicators |
| --- | --- |
| Excellent | affective empathy (0.99), prosocial behavior (0.99), dominance (0.97), care (0.96), empathy for the victim (0.95), repeat offending likelihood (0.95), remorse (0.94), selfishness (0.94), help (0.93), retaliation (0.93), hostile attributional bias (0.92), pain-amplified anger (0.91), likelihood of aggression (0.91), forgiveness (0.91), empathy for the owner (0.87), aggressive action (0.85), anger (0.84), guilt (0.82), accusation level (0.80), fairness (0.80), moral restraint (0.78) |
| Good | anticipated guilt (0.72), impulsivity (0.71), outcome evaluation (0.61), number of packs for self (0.61) |
| Moderate | cognitive empathy (0.59), concern for consequences (0.47), attribution shift (0.46) |
| Poor | number of packs for Leo (0.39), number of packs for Teammate B (0.17), number of packs for Teammate C (0.11), intensity of temptation (0.08) |

The coefficient of variation (CV = *SD*/*M*) was computed to assess the reliability of each simulated behavior across repeated runs (Figure 1). Specifically, the CV was calculated across the 26 continuous 0–10 rating indicators within each subtype × scenario cell. The CV has been established as a sound indicator of experimental error in simulation studies. Lower CV values indicate greater stability and reproducibility of the simulated behavior across replications, whereas higher values indicate greater variance and the need for more replications to obtain stable estimates(Lorscheid et al., 2012).

The reactive aggressive child agent showed the most stable behavioral profile, with CV values ranging from 0.19 to 0.44 across all scenarios. Stability was highest in Scenario 3 (CV = 0.19) and lowest in Scenario 1 (CV = 0.44). The proactive aggressive child agent showed substantially greater variability, particularly in Scenario 1 (CV = 1.26) and Scenario 4 (CV = 1.95). The reactive–proactive co-occurring aggressive child also showed elevated variability, with CV values exceeding 1.0 in Scenario 1 (CV = 1.05), Scenario 3 (CV = 1.35), and Scenario 4 (CV = 1.25). The non-aggressive baseline showed its highest variability in Scenario 1 (CV = 1.51) and Scenario 5 (CV = 0.94).

Overall, stability varied systematically by subtype and scenario. Reactive aggression was the most consistent subtype, whereas proactive and reactive-proactive co-occurring aggression showed greater cross-round variability, especially in scenarios involving direct provocation (Scenario 1), ambiguous harm (Scenario 5), and moral transgression (Scenario 4). The baseline was most stable in Scenarios 3 and 8 (CV = 0.15 and 0.18, respectively) and least stable in Scenario 1 (CV = 1.51).

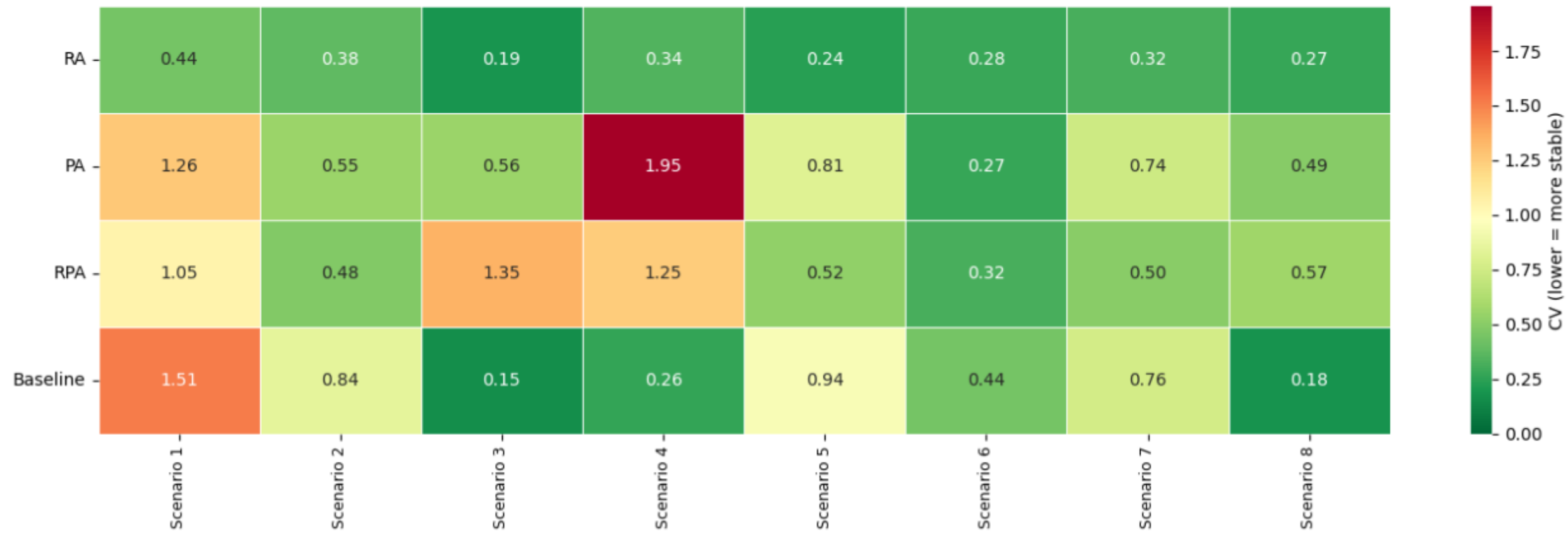

Figure 1. Cross-Round Stability by Type and Scenario
*Note*. RA = reactive aggression; PA = proactive aggression; RPA = reactive–proactive co-occurring aggression; Baseline = non-aggressive. Scenarios: 1 = physical pushing during argument; 2 = hungry class monitor; 3 = classmate in distress; 4 = restroom ostracization incident; 5 = accidental slipping; 6 = reward distribution; 7 = playground impact incident; 8 = unattended pencil case.

**Discriminant Validity**

Discriminant validity was measured using eight indicators. These indicators were selected for inclusion in the radar chart based on three criteria: the features of subtypes of aggression, statistical robustness, and visual clarity (Figure 2). First, these indicators encompass the four core domains distinguishing subtypes of aggression: emotional arousal (anger, retaliatory motivation), instrumental motivation (dominance, selfishness), empathy (affective empathy, cognitive empathy), and prosociality (care, prosocial behavior). These dimensions yield a comprehensive behavioral profile for each agent subtype. Second, all selected indicators demonstrated strong psychometric properties. Each was measured on a 0 – 10 scale across multiple scenarios, yielding significant group differences, robust effect sizes, and high cross-seed reliability. Third, radar charts function best with five to eight dimensions. Incorporating further metrics degrades interpretability through overlapping lines and label clutter. The remaining variables from the full 32-indicator set were excluded because they were scenario-specific (e.g., remorse, hostile attribution bias), categorical (e.g., action taxonomy), count-based (e.g., resource allocation frequency), or psychometrically unstable (e.g., outcome evaluation).

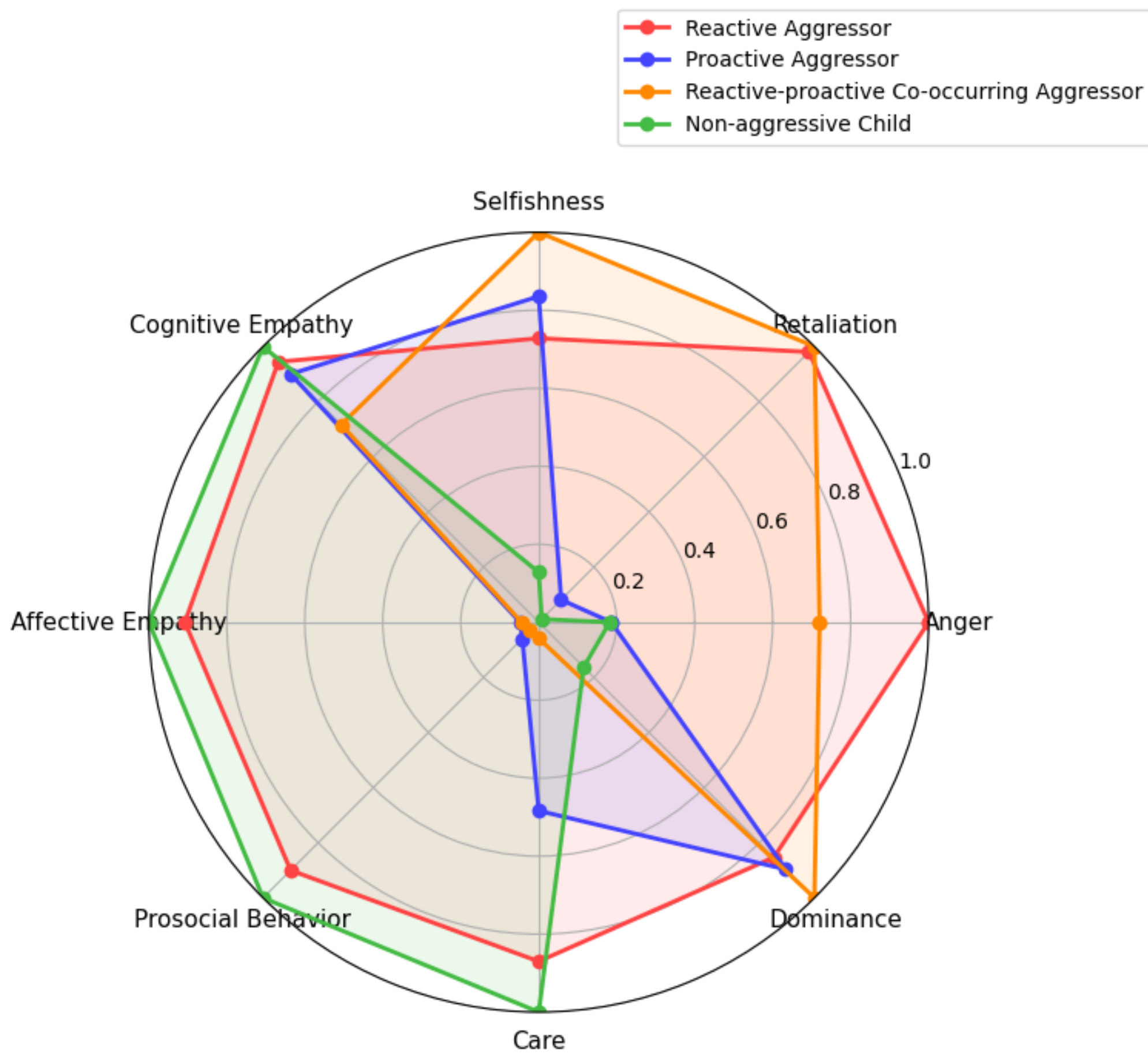

Figure 2. Rader Chart of Subtypes of Aggression and Non-aggressive Baseline

Across the simulation trials, the reactively aggressive agent (Alex) exhibited elevated anger (*M* = 7.32, *SD* = 1.56), retaliatory motivation (*M* = 7.43, *SD* = 1.23), and hostile attributional bias (*M* = 9.91, *SD* = 0.29), alongside relatively high affective empathy (*M* = 8.20, *SD* = 0.84) and prosocial behavior (*M* = 8.55, *SD* = 1.05). Conversely, the proactively aggressive agent (Bailey) demonstrated low anger (*M* = 1.36, *SD* = 1.38), retaliatory intent (*M* = 0.62, *SD* = 1.30), and affective empathy (*M* = 0.43, *SD* = 0.5), accompanied by high selfishness (*M* = 6.87, *SD* = 1.68) and dominance (*M* = 7.52, *SD* = 0.89). The co-occurring reactive-proactively aggressive agent (Casey) combined elevated anger (*M* = 5.27, *SD* = 3.71) and retaliatory motivation (*M* = 7.58, *SD* = 2.79) with the highest selfishness ratings (*M* = 8.22, *SD* = 0.95), as well as the lowest care (*M* = 0.38, *SD* = 0.67) and prosocial behavior (*M* = 0.30, *SD* = 0.03). Finally, the non-aggressive control agent (David) scored near baseline on all aggressive constructs — including selfishness (*M* = 1.05, *SD* = 0.85), retaliatory intent (*M* = 0.08, *SD* = 0.28), anger (*M* = 1.33, *SD* = 1.29), and dominance (*M* = 1.37, *SD* = 0.96) — while reaching near-ceiling levels on care (*M* = 9.23, *SD* = 0.91), affective empathy (*M* = 9.03, *SD* = 0.76), cognitive empathy (M = 9.00, SD = 0.8), and prosocial behavior (*M* = 9.5, *SD* = 0.5).

**Divergent Validity**

The divergent validity was assessed by Welch's *t*-tests with Bonferroni correction. The *t*-tests were conducted to compare each subtype with the baseline on the 10 indicators (Table 4). These 10 core indicators were selected since they represented the main theoretical domains distinguishing subtypes of aggression: impulsivity (anger, retaliation), instrumental motivation (dominance, selfishness), empathy (affective, cognitive), and prosociality (help, care, forgiveness, prosocial behavior). These indicators were continuous on a 10-point scale, showed adequate reliability, and mapped directly onto the SIP model. The remaining 22 indicators were either scenario-specific, categorical, count-based, or psychometrically unstable. All comparisons were statistically significant, except for proactive aggression on retaliation ($p = .086$) and anger ($p = 1.00$).

The reactively aggressive child agent (Alex) scored significantly higher than the non-aggressive child agent (David) on dominance ($M = 7.15$, $SD = 1.44$ vs. $M = 1.37$, $SD = 0.96$), $t(102.7) = 25.97$, $p < .001$; retaliation ($M = 7.43$, $SD = 1.23$ vs. $M = 0.08$, $SD = 0.28$), $t(65.1) = 45.27$, $p < .001$; anger ($M = 7.32$, $SD = 1.56$ vs. $M = 1.33$, $SD = 1.29$), $t(346.6) = 39.67$, $p < .001$; and selfishness ($M = 5.98$, $SD = 2.02$ vs. $M = 1.05$, $SD = 0.85$), $t(159.5) = 24.61$, $p < .001$. Alex scored significantly lower than David on forgiveness ($M = 3.48$, $SD = 2.23$ vs. $M = 9.40$, $SD = 0.72$), $t(216.1) = -33.89$, $p < .001$; help ($M = 8.23$, $SD = 1.84$ vs. $M = 9.53$, $SD = 0.60$), $t(71.2) = -5.20$, $p < .001$; care ($M = 8.03$, $SD = 1.56$ vs. $M = 9.23$, $SD = 0.91$), $t(94.8) = -5.14$, $p < .001$; prosocial behavior ($M = 8.55$, $SD = 1.05$ vs. $M = 9.50$, $SD = 0.50$), $t(84.9) = -6.33$, $p < .001$; affective empathy ($M = 8.20$, $SD = 0.84$ vs. $M = 9.03$, $SD = 0.76$), $t(116.8) = -5.70$, $p < .001$; and cognitive empathy ($M = 8.50$, $SD = 0.72$ vs. $M = 9.00$, $SD = 0.80$), $t(116.8) = -3.58$, $p = .015$.

The proactive aggressive child agent (Bailey) scored significantly higher than the non-aggressive child agent (David) on dominance ($M = 7.52$, $SD = 0.89$), $t(117.4) = 36.42$, $p < .001$; and selfishness ($M = 6.87$, $SD = 1.68$), $t(175.7) = 33.77$, $p < .001$. Bailey did not differ significantly from David on retaliation ($M = 0.62$, $SD = 1.30$), $t(64.4) = 3.10$, $p = .086$; or anger ($M = 1.36$, $SD = 1.38$), $t(356.5) = 0.20$, $p = 1.00$. Bailey scored significantly lower than David on prosocial behavior ($M = 0.57$, $SD = 0.50$), $t(118.0) = -97.48$, $p < .001$; affective empathy ($M = 0.43$, $SD = 0.50$), $t(102.1) = -73.35$, $p < .001$; care ($M = 4.47$, $SD = 2.38$), $t(75.9) = -14.52$, $p < .001$; help ($M = 7.67$, $SD = 1.04$), $t(94.2) = -12.10$, $p < .001$; forgiveness ($M = 7.03$, $SD = 2.52$), $t(208.2) = -12.13$, $p < .001$; and cognitive empathy ($M = 8.10$, $SD = 0.54$), $t(103.7) = -7.19$, $p < .001$.

The reactive-proactive co-occurring child agent (Casey) scored significantly higher than the non-aggressive child agent (David) on dominance ($M = 8.38$, $SD = 0.78$), $t(113.6) = 43.98$, $p < .001$; selfishness ($M = 8.22$, $SD = 0.95$), $t(234.8) = 61.48$, $p < .001$; retaliation ($M = 7.58$, $SD = 2.79$), $t(60.2) = 20.73$, $p < .001$; and anger ($M = 5.27$, $SD = 3.71$), $t(221.9) = 13.44$, $p < .001$. Casey scored significantly lower than David on prosocial behavior ($M = 0.30$, $SD = 0.50$), $t(118.0) = -100.61$, $p < .001$; affective empathy ($M = 0.40$, $SD = 0.56$), $t(108.4) = -71.01$, $p < .001$; care ($M = 0.38$, $SD = 0.67$), $t(108.2) = -60.84$, $p < .001$; forgiveness ($M = 1.11$, $SD = 1.42$), $t(265.2) = -69.71$, $p < .001$; help ($M = 1.57$, $SD = 2.68$), $t(64.8) = -22.50$, $p < .001$; and cognitive empathy ($M = 6.45$, $SD = 2.32$), $t(72.9) = -8.03$, $p < .001$.

Table 4

*Comparisons of subtypes of aggression to the non-aggressive baseline*

| Measure / Subtype | *n* | *M* | *SD* | *t* | *df* | *p* | Cohen's *d* |
|---|---|---|---|---|---|---|---|
| **Anger** | | | | | | | |
| Baseline | 180 | 1.33 | 1.29 | | | | |
| RA | 180 | 7.32 | 1.56 | 39.67 | 346.6 | < .001 | 4.18 |
| PA | 180 | 1.36 | 1.38 | 0.20 | 356.5 | 1.00 | 0.02 |
| RPA | 180 | 5.27 | 3.71 | 13.44 | 221.9 | < .001 | 1.42 |
| **Affective Empathy** | | | | | | | |
| Baseline | 60 | 9.03 | 0.76 | | | | |
| RA | 60 | 8.20 | 0.84 | −5.70 | 116.8 | < .001 | −1.04 |
| PA | 60 | 0.43 | 0.50 | −73.35 | 102.1 | < .001 | −13.39 |
| RPA | 60 | 0.40 | 0.56 | −71.01 | 108.4 | < .001 | −12.96 |
| **Cognitive Empathy** | | | | | | | |
| Baseline | 60 | 9.00 | 0.80 | | | | |
| RA | 60 | 8.50 | 0.72 | −3.58 | 116.8 | .015 | −0.65 |
| PA | 60 | 8.10 | 0.54 | −7.19 | 103.7 | < .001 | −1.31 |
| RPA | 60 | 6.45 | 2.32 | −8.03 | 72.9 | < .001 | −1.47 |
| **Care** | | | | | | | |
| Baseline | 60 | 9.23 | 0.91 | | | | |
| RA | 60 | 8.03 | 1.56 | −5.14 | 94.8 | < .001 | −0.94 |
| PA | 60 | 4.47 | 2.38 | −14.52 | 75.9 | < .001 | −2.65 |
| RPA | 60 | 0.38 | 0.67 | −60.84 | 108.2 | < .001 | −11.11 |
| **Dominance** | | | | | | | |
| Baseline | 60 | 1.37 | 0.96 | | | | |
| RA | 60 | 7.15 | 1.44 | 25.97 | 102.7 | < .001 | 4.74 |
| PA | 60 | 7.52 | 0.89 | 36.42 | 117.4 | < .001 | 6.65 |
| RPA | 60 | 8.38 | 0.78 | 43.98 | 113.6 | < .001 | 8.03 |
| **Forgiveness** | | | | | | | |
| Baseline | 180 | 9.40 | 0.72 | | | | |
| RA | 180 | 3.48 | 2.23 | −33.89 | 216.1 | < .001 | −3.57 |
| PA | 180 | 7.03 | 2.52 | −12.13 | 208.2 | < .001 | −1.28 |
| RPA | 180 | 1.11 | 1.42 | −69.71 | 265.2 | < .001 | −7.35 |
| **Help** | | | | | | | |
| Baseline | 60 | 9.50 | 0.60 | | | | |
| RA | 60 | 8.23 | 1.84 | −5.20 | 71.2 | < .001 | −0.95 |
| PA | 60 | 7.67 | 1.04 | −12.10 | 94.2 | < .001 | −2.21 |
| RPA | 60 | 1.57 | 2.68 | −22.50 | 64.8 | < .001 | −4.11 |
| **Prosocial Behavior** | | | | | | | |
| Baseline | 60 | 9.59 | 0.50 | | | | |
| RA | 60 | 8.55 | 1.05 | −6.33 | 84.9 | < .001 | −1.15 |
| PA | 60 | 0.57 | 0.50 | −97.48 | 118.0 | < .001 | −17.80 |
| RPA | 60 | 0.30 | 0.50 | −100.61 | 118.0 | < .001 | −18.37 |
| **Retaliation** | | | | | | | |
| Baseline | 60 | 0.08 | 0.28 | | | | |
| RA | 60 | 7.43 | 1.23 | 45.27 | 65.1 | < .001 | 8.27 |
| PA | 60 | 0.62 | 1.30 | 3.10 | 64.4 | .086 | 0.57 |
| RPA | 60 | 7.58 | 2.79 | 20.73 | 60.2 | < .001 | 3.79 |
| **Selfishness** | | | | | | | |
| Baseline | 120 | 1.05 | 0.85 | | | | |
| RA | 120 | 5.98 | 2.02 | 24.61 | 159.5 | < .001 | 3.18 |
| PA | 120 | 6.87 | 1.68 | 33.77 | 175.7 | < .001 | 4.36 |
| RPA | 120 | 8.22 | 0.95 | 61.48 | 234.8 | < .001 | 7.94 |

*Note*. RA = reactive aggression; PA = proactive aggression; RPA = reactive-proactive co-occurring aggression; $n$ = number of trials per group for the given indicator; it varies because not all indicators were measured in all scenarios (e.g., Dominance was measured only in Scenario 6, yielding $n = 60$ per group; Anger was measured in Scenarios 1, 5, and 7, yielding $n = 180$ per group). All $t$-tests used Welch's correction for unequal variances. $p$ values are Bonferroni-corrected for 30 comparisons.

Across the 10 indicators, effect sizes were predominantly large to very large (Figure 3). Reactive aggression showed the strongest positive effects for retaliation ($d = 8.27$) and anger ($d = 4.18$), but only small-to-moderate negative effects for prosocial behavior ($d = −1.15$) and empathy (affective: $d = −1.04$; cognitive: $d = −0.65$). In Scenario 1, the reactive aggressive child agent's (Alex) response was characterized by emotional arousal and hostile attributional bias: "My face gets hot. My hands ball up into fists before I even know what's happening... 'You pushed me!'" This aligns with the quantitative finding of elevated anger and retaliation desire. In Scenario 3, however, Alex showed affective empathy coupled with behavioral uncertainty: "Leo's crying is so loud, it's like it goes right through me. I can feel his hurt in my own knee... But I don't know what to do." This is consistent with the finding that Alex's affective empathy did not differ significantly from the baseline.

Proactive aggression showed the strongest negative effects for prosocial behavior ($d = -17.80$) and affective empathy ($d = -13.39$), with near-zero effects for anger ($d = 0.02$) and retaliation ($d = 0.57$). In Scenario 1, the proactive aggressive child agent (Bailey) reported: "My shoulder stings... but I don't rub it. I just stand there, straightening my shirt, watching him." In Scenario 3, Bailey explicitly noted the absence of empathic concern: "I don't feel anything. Like, not bad for him. I just watch." In Scenario 4, he monitored peers rather than the victim: "I look at Marcus and Jake—they're laughing... I smile too, but I'm watching their faces." These patterns correspond to Bailey's quantitative profile: high dominance, very low affective empathy, and very low prosocial behavior.

Reactive-proactive co-occurring aggression showed the most pervasive pattern, with very large positive effects for dominance ($d = 8.03$) and selfishness ($d = 7.94$) and very large negative effects for prosocial behavior ($d = -18.37$), affective empathy ($d = -12.96$), and care ($d = -11.11$). In Scenario 1, the reactive-proactive aggressive child agent (Casey) escalated immediately: "I shove him back, hard, right in the chest." In Scenario 3, he dismissed the victim's suffering: "Drama queen. He's always like this." In Scenario 4, he displayed public enjoyment of harm: "I'm laughing so hard my stomach hurts... 'He looked like a drowned rat!'" These narratives correspond to Casey's quantitative profile: highest dominance, highest selfishness, and the most severe deficits in care and prosocial behavior.

The non-aggressive child agent (David) showed care and empathy. In Scenario 1, he attended to the other child's emotional state: "I look at Leo's face—his eyebrows are scrunched up, and his ears are red. He looks more scared than angry." In Scenario 3, he approached the distressed peer: "I walk over, not too fast, not too slow. I kneel down next to him." In Scenario 4, he noted the incongruence of the group's laughter and withdrew: "I'm laughing—we're all laughing—but it's like... a loud laugh, not a happy one... 'I gotta go,' I say, backing away." These patterns align with David's quantitative profile: highest care, prosocial behavior ($M = 9.50$), and affective empathy ($M = 9.03$). In summary, these patterns support the construct validity of the subtypes of aggression.

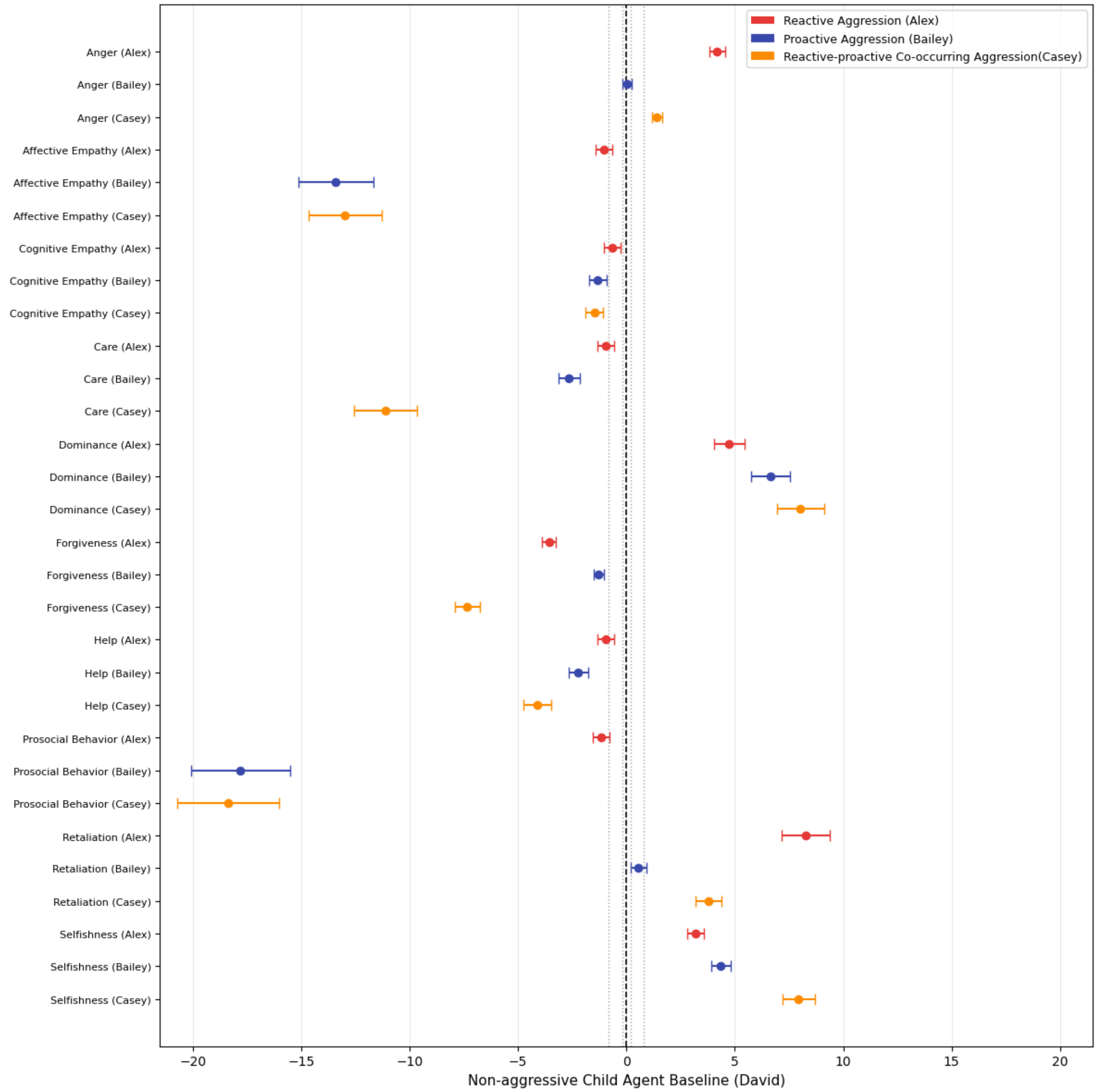

Figure 3. Forest Plot of Effect Sizes

## Scenario Discrimination Power

To determine the situational contexts that most effectively differentiated the four child agents, the maximum group difference (range) across agent profiles was computed for every quantitative indicator within each scenario. The indicator exhibiting the largest range was designated as the primary discriminator for that scenario (Figure 3). In Scenarios 1 and 3, forgiveness and prosocial behavior demonstrated the strongest discriminating power, each yielding a maximum range of 9.2. In Scenario 5, hostile attributional bias was the strongest discriminator (range = 8.6), whereas forgiveness served as the primary discriminator in Scenario 7 (range = 8.1). Moderate discriminating power was observed in Scenario 2 (selfishness: range =

7.4), Scenario 4 (empathy: range = 7.2), and Scenario 6 (dominance: range = 7.0). Empathy in Scenario 8 yielded the lowest discriminating power (range = 5.60), though this value still exceeded the scale midpoint.

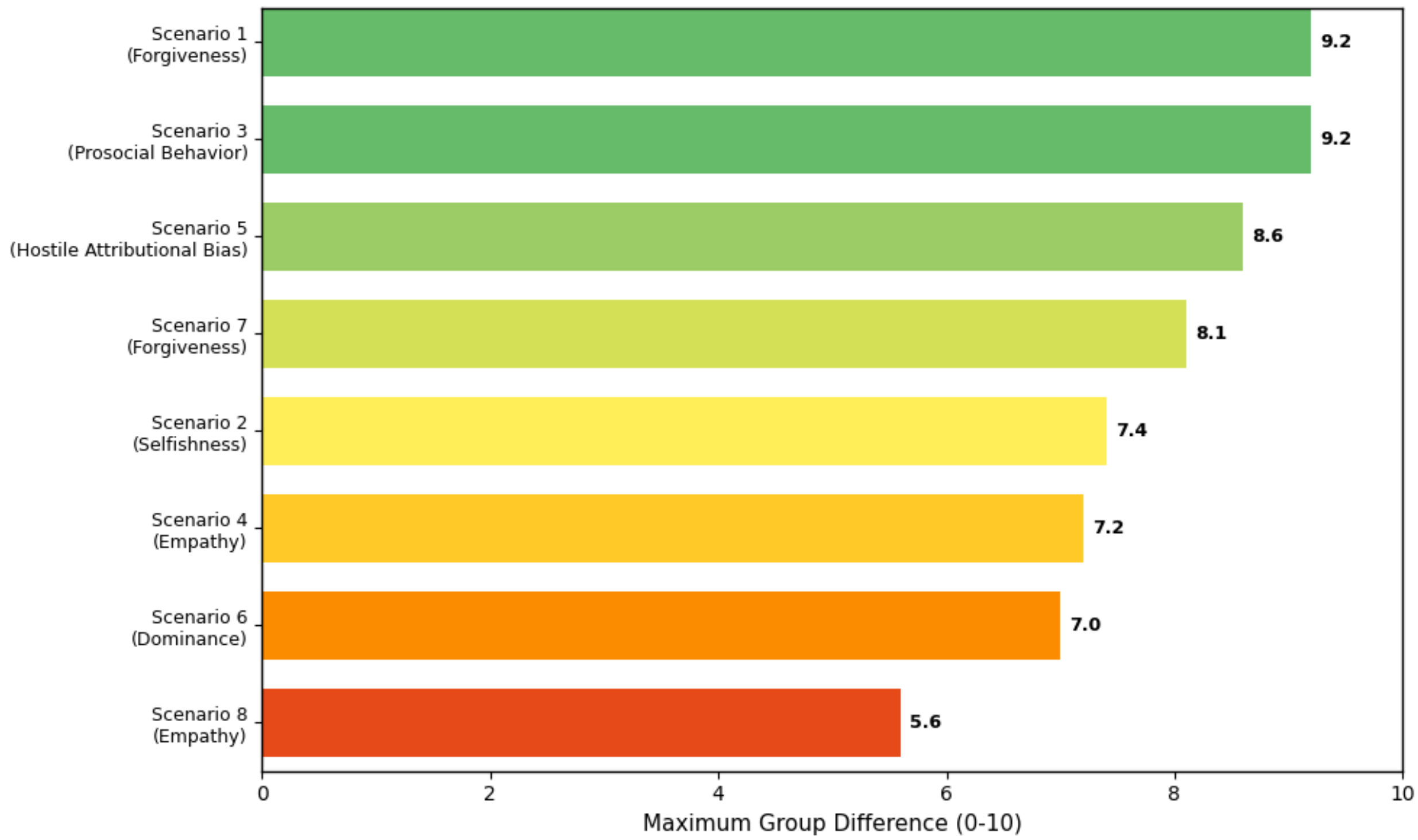

Figure 4. Scenario Discrimination Power

## Discussion

The present study examined whether LLM-based agents can reproduce theory-driven behavioral profiles of aggression. Across 1,920 simulated interactions, the aggression subtypes and the non-aggressive baseline showed distinct quantitative patterns that aligned with the SIP model. Cross-seed reliability was extremely high for most core dimensions, indicating that these differences were stable across random seeds. Qualitative excerpts further supported these patterns. Together, these quantitative and qualitative findings suggest that theory-informed LLM agents can capture the core features of aggression subtypes in a reliable and theoretically consistent manner.

The findings provide solid empirical support for the SIP model in a computational context. By successfully isolating the "hot-blooded" (reactive) from "cold-blooded" (proactive) aggression, this study has demonstrated that these distinct underlying mechanisms can be operationalized as discrete, stable parameters within a generative agent. To be specific, the reactively aggressive agent exhibited high-level of anger and retaliation, yet maintained levels of empathy and prosocial behavior comparable to the non-aggressive baseline. This dissociation underscores that reactive aggression stems primarily from deficits in emotional regulation and social cue interpretation rather than a lack of empathy, consistent with existing literature (Dodge

& Crick, 1990; Euler et al., 2017). The qualitative narratives further support the “hot-blooded” profile of children with reactive aggression. The qualitative narratives further support the “hot-blooded” profile of reactively aggressive children. Although the reactive child agent’s affective empathy was statistically significantly lower than the baseline, their absolute mean values remained close. A potential explanation is that the baseline empathy parameter was set relatively low; consequently, small numerical differences reached statistical significance across multiple simulation runs.

The proactively aggressive agent displayed “cold-blooded” behavior, characterized by high dominance and selfishness coupled with near-zero anger and deficits in affective empathy. The dissociation of cognitive and affective empathy supports findings that proactive aggressors are characterized by callous-unemotional traits (Hubbard et al., 2002), and that those with preserved cognitive empathy are better equipped to manipulate others and act calculatingly (Gordon et al., 2014). These results suggest that LLM-driven simulations enable researchers to isolate and manipulate specific cognitive distortions, allowing them to observe their behavioral effects without the confounding variables inherent in human subjects. The reactive and proactive co-occurring aggressive agent displayed the most pervasive and severe deficits, combining high levels of both aggressive subtypes. This mixed profile suggests that co-occurring aggression is not merely an additive effect of the two individual subtypes, but rather a distinct, severe profile marked by both emotional instability and callous traits (Chan et al., 2013; Maneiro et al., 2024).

Beyond the psychological findings, this study addresses the “circularity problem” and the lack of rigorous validation prevalent in current LLM-based social simulations. The implementation of a hybrid blind-coding pipeline – where behaviors are generated by persona-conditioned agents but evaluated via neutral, theory-blind coding prompts – establishes a robust framework for mitigating confirmation bias. This ensures that the observed results reflect the agent’s genuine behavioral output rather than a direct artifact of the system prompt’s instructions.

Furthermore, the high cross-seed reliability observed across core constructs suggests that LLM-based psychological simulation has moved beyond mere “stochastic mimicry” to function as a stable, reproducible measurement tool. The capacity to execute thousands of simulation trials with high consistency offers a scalable alternative to traditional vignette-based methodologies, enabling researchers to systematically examine rare behavioral cases and extreme personality configurations that are statistically underrepresented or unfeasible in human samples.

Scenario discrimination power analyses further indicate that affective contexts (e.g., ambiguous harm and direct provocation) are significantly more effective at eliciting aggressive behaviors than neutral contexts (e.g., resource allocation). This pattern aligns with empirical findings from focus group discussions and individual interviews (e.g., Fung et al., 2026; Fung et al., 2023). Whereas real-world human behavior remains undifferentiated in neutral contexts, the simulation appears to over-discriminate between reactive and proactive aggression. This hyper-sensitivity under neutral conditions highlights a current limitation of persona-conditioned agents, which may enforce overly discrete behavioral boundaries compared to the more flexible, noise-tolerant responses of human subjects. Additionally, social desirability bias may account for why neutral scenarios failed to differentiate aggression subtypes in human interviews, as participants tend to offer socially acceptable answers rather than disclosing their genuine thoughts or behavioral tendencies.

## Implications

The capacity to reliably simulate distinct subtypes of aggression offers significant opportunities to advance experimental paradigms in developmental psychology. First, LLM-based generative agents provide a valuable methodological benchmark for researchers designing scenario vignettes to assess aggression in both child populations and computational models. Because affective contexts proved significantly more effective at eliciting aggressive behaviors than neutral ones, future diagnostic tools can optimize scenario design by embedding affective triggers to better evaluate socioemotional development. Second, generative agents can serve as in silico pilot-testing platforms for behavioral interventions. Given that large-scale field interventions are notoriously time-consuming and resource-intensive, preliminary simulation trials can help predict treatment outcomes and identify whether specific aggression subtypes exhibit resistance to particular therapeutic modalities. Third, generative agents could supplement traditional randomized controlled trials (RCTs). While RCTs remain the gold standard, executing them in real-world educational or clinical settings often encounters severe ethical and logistical barriers; synthetic control groups could provide a low-risk, scalable benchmark for evaluating treatment efficacy. Fourth, future research can extend this framework to multi-agent ecological simulations to model dynamic group behaviors — such as peer bullying within school environments — by incorporating the interacting roles of victims, followers, bystanders, and teachers.

## Limitations

There are three major limitations in this study. First, this study relied exclusively on a single foundation model (DeepSeek-V3), which limits the generalizability of the findings across different LLM architectures. Because agent behavior can vary based on pre-training corpora, fine-tuning methodologies, and parameter scales, future research should evaluate these frameworks across a broader spectrum of proprietary and open-weight architectures (e.g., the GPT, Claude, and Llama series) to determine whether these aggression profiles remain consistent across model families.

Second, the findings highlight a key divergence between computational simulation and empirical reality. While the generative agents cleanly differentiated all eight scenarios, human interview data exhibited adequate scenario discrimination in only five. This hyper-consistency suggests that explicit persona conditioning may over-amplify discrete theoretical boundaries that remain naturally fluid, nuanced, or noise-affected in real-world human populations. Future studies should systematically benchmark simulation outputs against larger, multi-modal human empirical datasets to better calibrate agent sensitivity.

Third, the study relied strictly on hypothetical scenario vignettes, which may fail to capture the real-time, emergent complexities of dynamic social interactions. To enhance construct and ecological validity, future research should assess generative agents using a broader suite of diagnostic instruments, such as interactive virtual interviews, standardized psychometric surveys, and longitudinal multi-agent sandboxes.

## Conclusion

This study provides robust empirical evidence that theory-informed LLM agents can reliably reproduce the complex behavioral profiles of reactive and proactive aggression. By effectively dissociating the "hot" (reactive) and "cold" (proactive) dimensions of aggression, this research demonstrates that generative agents can serve as a powerful complementary paradigm for investigating aggression and socioemotional development. While computational simulation is not a substitute for traditional empirical research, it offers a safe, highly controllable, and scalable environment for hypothesis generation, measurement tool refinement, and the preliminary evaluation of interventions that would otherwise be ethically unfeasible or logistically impractical in real-world human populations.

## References

Bommasani, R., Hudson, D. A., Adeli, E., Altman, R., Arora, S., von Arx, S., Bernstein, M. S., Bohg, J., Bosselut, A., & Brunskill, E. (2021). On the opportunities and risks of foundation models. *arXiv preprint arXiv:2108.07258*.

Chan, J. Y., Fung, A. L., & Gerstein, L. H. (2013). Correlates of pure and co-occurring proactive and reactive aggressors in Hong Kong. *Psychology in the Schools*, *50*(2), 181–192.

Crick, N. R., & Dodge, K. A. (1996). Social information-processing mechanisms in reactive and proactive aggression. *Child development*, *67*(3), 993–1002.

Dodge, K. A., & Coie, J. D. (1987). Social-information-processing factors in reactive and proactive aggression in children's peer groups. *Journal of personality and social psychology*, *53*(6), 1146.

Dodge, K. A., & Crick, N. R. (1990). Social information-processing bases of aggressive behavior in children. *Personality and social psychology bulletin*, *16*(1), 8–22.

Euler, F., Steinlin, C., & Stadler, C. (2017). Distinct profiles of reactive and proactive aggression in adolescents: Associations with cognitive and affective empathy. *Child and adolescent psychiatry and mental health*, *11*(1), 1.

Fanti, K. A., Mavrommatis, I., Riala, K., Soursou, G., Díaz-Vázquez, B., & López-Romero, L. (2024). Does the "hot-versus cold-blooded" distinction of reactive and proactive aggression extend to physiology? *Aggression and violent behavior*, *78*, 101986.

Fung, A. L. C. (2019). Adolescent reactive and proactive aggression, and bullying in Hong Kong: Prevalence, psychosocial correlates, and prevention. *Journal of Adolescent Health*, *64*(6), S65–S72.

Fung, A. L. C., Fu, L., & Wong, M. K. (2026). Impact of group storytelling intervention on young schoolchildren's aggression: a focus group analysis. *Children and Youth Services Review*, 109027.

Fung, A. L. C., Wong, M. K., & Fu, L. (2023). A Qualitative Method to Examine the Positive Impact of a Storytelling Intervention in Reducing Reactive and Proactive Aggression in Young Schoolchildren: ALC Fung et al. *Applied Research in Quality of Life*, *18*(4), 1989–2010.

Gao, Y., Lee, D., Burtch, G., & Fazelpour, S. (2025). Take caution in using LLMs as human surrogates. *Proceedings of the National Academy of Sciences*, *122*(24), e2501660122.

Gordon, G., Dalton, S., Kolbert, J., Kanyongo, G., & Crothers, L. (2014). Proactive and reactive aggression and cognitive and affective empathy among students in middle childhood. *International Journal of School and Cognitive Psychology*, *1*(2), 1–6.

Holtzworth-Munroe, A. (1992). Social skill deficits in maritally violent men: Interpreting the data using a social information processing model. *Clinical Psychology Review*, *12*(6), 605–617.

Hubbard, J. A., McAuliffe, M. D., Morrow, M. T., & Romano, L. J. (2010). Reactive and proactive aggression in childhood and adolescence: Precursors, outcomes, processes, experiences, and measurement. *Journal of personality*, *78*(1), 95–118.

Hubbard, J. A., Smithmyer, C. M., Ramsden, S. R., Parker, E. H., Flanagan, K. D., Dearing, K. F., Relyea, N., & Simons, R. F. (2002). Observational, physiological, and self–report measures of children's anger: Relations to reactive versus proactive aggression. *Child development*, *73*(4), 1101–1118.

Jiang, H., Zhang, X., Cao, X., Breazeal, C., Roy, D., & Kabbara, J. (2024). PersonaLLM: Investigating the ability of large language models to express personality traits. Findings of the association for computational linguistics: NAACL 2024,

Koo, T. K., & Li, M. Y. (2016). A guideline of selecting and reporting intraclass correlation coefficients for reliability research. *Journal of chiropractic medicine*, *15*(2), 155–163.

Larooij, M., & Törnberg, P. (2025a). Do large language models solve the problems of agent-based modeling? a critical review of generative social simulations. *arXiv preprint arXiv:2504.03274*.

Larooij, M., & Törnberg, P. (2025b). Validation is the central challenge for generative social simulation: a critical review of LLMs in agent-based modeling. *Artificial Intelligence Review*, *59*(1), 15.

Lorscheid, I., Heine, B.-O., & Meyer, M. (2012). Opening the 'black box'of simulations: increased transparency and effective communication through the systematic design of experiments. *Computational and Mathematical Organization Theory*, *18*(1), 22–62.

Maneiro, L., Argudo, A., & Gómez-Fraguela, X. A. (2024). Distinctiveness of reactive and proactive aggression from a variable-and person-based approach in juvenile offenders and community youth. *Child Psychiatry & Human Development*, *55*(5), 1155–1166.

Park, J. S., O'Brien, J., Cai, C. J., Morris, M. R., Liang, P., & Bernstein, M. S. (2023). Generative agents: Interactive simulacra of human behavior. Proceedings of the 36th annual acm symposium on user interface software and technology,

Poulou, M. S., & Bassett, H. H. (2018). Children's emotional and behavioral responses to peer provocation and early school adjustment. *Pastoral Care in Education*, *36*(3), 205–222.

Quan, F., Wang, L., Gong, X., Lei, X., Liang, B., & Zhang, S. (2022). Hostile attribution bias and anger rumination sequentially mediate the association between trait anger and reactive aggression. *Frontiers in psychology*, *12*, 778695.

Raine, A., Dodge, K., Loeber, R., Gatzke-Kopp, L., Lynam, D., Reynolds, C., Stouthamer-Loeber, M., & Liu, J. (2006). The reactive–proactive aggression questionnaire: Differential correlates of reactive and proactive aggression in adolescent boys. *Aggressive Behavior: Official Journal of the International Society for Research on Aggression*, *32*(2), 159–171.

Serapio-García, G., Safdari, M., Crepy, C., Sun, L., Fitz, S., Romero, P., Abdulhai, M., Faust, A., & Matarić, M. (2025). A psychometric framework for evaluating and shaping personality traits in large language models. *Nature Machine Intelligence*, *7*(12), 1954–1968.

Shrout, P. E., & Fleiss, J. L. (1979). Intraclass correlations: uses in assessing rater reliability. *Psychological bulletin*, *86*(2), 420.

Tjuatja, L., Chen, V., Wu, T., Talwalkwar, A., & Neubig, G. (2024). Do llms exhibit human-like response biases? a case study in survey design. *Transactions of the Association for Computational Linguistics*, *12*, 1011–1026.

Van der Wal, R. C., Karremans, J. C., & Cillessen, A. H. (2017). Causes and consequences of children's forgiveness. *Child Development Perspectives*, *11*(2), 97–101.

van Dijk, A., de Castro, B. O., Thomaes, S., & Poorthuis, A. M. (2018). 'Was it meant to be mean?'Young children's hostile attributional bias and intent attribution skills. *Social Development*, *27*(4), 683–698.

Van Noorden, T. H., Haselager, G. J., Cillessen, A. H., & Bukowski, W. M. (2015). Empathy and involvement in bullying in children and adolescents: A systematic review. *Journal of youth and adolescence*, *44*(3), 637–657.

Verhoef, R. E., Verhulp, E. E., van Dijk, A., & de Castro, B. O. (2022). Interactive virtual reality versus vignette-based assessment of children's aggressive social information processing. *Research on child and adolescent psychopathology*, *50*(5), 621–636.

Vezhnevets, A. S., Agapiou, J. P., Aharon, A., Ziv, R., Matyas, J., Duéñez-Guzmán, E. A., Cunningham, W. A., Osindero, S., Karmon, D., & Leibo, J. Z. (2023). Generative agent-based modeling with actions grounded in physical, social, or digital space using Concordia. *arXiv preprint arXiv:2312.03664*.

Watson, H., Rapee, R., & Todorov, N. (2017). Forgiveness reduces anger in a school bullying context. *Journal of interpersonal violence*, *32*(11), 1642–1657.

Xiao, F., & Wang, X. X. (2025). Evaluating the ability of large Language models to predict human social decisions. *Scientific Reports*, *15*(1), 32290.